\documentclass[%
 reprint,
nofootinbib,
 amsmath,amssymb,
 aps,
prl,
showkeys,
floatfix,
]{revtex4-2}

\usepackage{graphicx}% Include figure files
\usepackage{dcolumn}% Align table columns on decimal point
\usepackage{bm}% bold math
\usepackage{hyperref}% add hypertext 
\usepackage{color}
\usepackage{xcolor}

\begin{document}

%\preprint{APS/123-QED}

\title{Hawking evaporation and the Landauer Principle}% Force line breaks with \\

\author{Marina Cort\^es}
%\affiliation{Instituto de Astrof\'{\i}sica e Ci\^{e}ncias do Espa\c{c}o, Universidade de Lisboa,
%Faculdade de Ci\^{e}ncias, Campo Grande, PT1749-016 Lisboa, Portugal}
\author{Andrew R.\ Liddle}
\affiliation{Institute of Astrophysics and Space Sciences, University of Lisbon, Faculty of Sciences,
%\affiliation{Instituto de Astrof\'{\i}sica e Ci\^{e}ncias do Espa\c{c}o, Universidade de Lisboa,
%Faculdade de Ci\^{e}ncias, 
Campo Grande, PT1749-016 Lisbon, Portugal}

\date{\today}% It is always \today, today,

\begin{abstract}
We show that Hawking black-hole evaporation saturates the Landauer Principle of information thermodynamics. Our result implies that information loss experienced by a black hole during Hawking evaporation takes place as efficiently as possible. We also extend the analysis to the case of Barrow entropy as a phenomenological realization of a fractal event horizon, where the Landauer Principle informs amongst different options for the black hole temperature. To our knowledge, this work is the first identification of the two results as  expressions of the same process.
\end{abstract}

%Use showkeys class option if keyword
\maketitle

%\tableofcontents

\section{Introduction}

An important 
%key 
result in algorithmic information theory is the Landauer Principle \cite{Landauer}, which states that the deletion of one bit of information necessarily dissipates an energy of at least $\ln 2 \, k_{\rm B} T$, where $k_{\rm B}$ is the Boltzmann constant and $T$ is the temperature of the ambient medium into which the energy is dissipated. While the result is not entirely uncontroversial (see e.g.\ Ref.~\cite{Norton} and references therein), it has a strong theoretical backing \cite{Bennett,Plenio,Sagawa,Frank} and some experimental support \cite{Jun,Berut}. It potentially provides an ultimate limit on the energy cost of computation, though modern computers operate many orders of magnitude above it.

Information also plays an important role in the understanding of black hole thermodynamics --- see Refs.~\cite{Almheiri,Raju} for recent reviews.  However, only very few works have directly related black hole thermodynamics to the Landauer Principle. For instance the Landauer Principle is not mentioned at all in Refs.~\cite{Almheiri,Raju}, while Herrera's review of the Landauer Principle in General Relativity \cite{Herrera} does not discuss black hole thermodynamics. Existing discussions, summarised later, are always in the context of information lost to the surroundings when accreted into a black hole.

In this article we take another route
and instead apply Landauer's Principle to the process of Hawking evaporation \cite{Hawking}, which reduces the information content of the black hole as measured by the Bekenstein--Hawking entropy formula \cite{Bekenstein,Hawking}. We find that Hawking evaporation exactly saturates the Landauer Principle. That is to say, the information loss to a black hole during its evaporation takes place as efficiently as possible. 

\section{Entropy, information, and black hole evaporation}

The black hole entropy formula, introduced by Bekenstein \cite{Bekenstein}, states that its entropy (a measure of the information contained in the black hole) is proportional to the surface area, with proportionality constant equal to $1/4$ when the surface area is measured in Planck units and the entropy in Boltzmann units. That is,
\begin{equation}
\label{e:BHent}
\frac{S_{\rm BH}}{k_{\rm B}} = \frac{c^3 }{G \hbar} \, \pi r_{\rm S}^2  = \frac{4 \pi G}{c \hbar} \,  M^2 \,,
\end{equation}
where $M$ and $r_{\rm S} = 2GM/c^2$ are the mass and Schwarzschild radius of the black hole and the others are the usual fundamental physical constants (Newton's gravitational constant $G$, the speed of light $c$, and the reduced Planck constant $\hbar$). 

Consistency of Eq.~\ref{e:BHent} with the laws of thermodynamics led Hawking to discover that black holes must radiate as a black-body, with the temperature given by the formula \cite{Hawking}
\begin{equation}
\label{e:TBH}
k_{\rm B} T_{\rm BH} = \frac{\hbar c^3}{8\pi G M} \,.
\end{equation}

Our goal here is to consider how efficiently the black hole loses its information during the radiation process. Consider the black hole mass loss sufficient to reduce its information content by one bit of information. Since
\begin{equation}
\frac{\Delta S_{\rm BH}}{S_{\rm BH}} = 2 \frac{\Delta M}{M} \,,
\end{equation}
the reduction in mass when losing one Boltzmann unit of entropy is
\begin{equation}
\Delta M =\frac{k_{\rm B} M}{2S_{\rm BH}} =  \frac{c \hbar}{8\pi G M} \,.
\end{equation}
Using the $\ln 2$ conversion between number of bits and entropy in Boltzmann units, the mass loss when reducing the information by one bit is
\begin{equation}
\Delta M = \ln 2 \, \frac{k_{\rm B} T_{\rm BH}}{c^2} \,.
\end{equation}
Hence, the energy dissipated to the surroundings is
\begin{equation}
\label{e:BHLandauer}
    \Delta E = \ln 2 \, k_{\rm B} T_{\rm BH} \,.
\end{equation}

To interpret this, we now must pay attention to the environment into which the black hole is radiating. It need not be to the vacuum; in principle the black hole could be radiating into an ambient medium at any temperature up to the Hawking temperature itself. If the dissipation had been less than Eq.~(\ref{e:BHLandauer}), say by a factor $\alpha$ (with $0 < \alpha < 1$), the black hole would have been able to radiate in a Landauer-violating fashion into a heat bath at a temperature between $\alpha T_{\rm BH}$ and $T_{\rm BH}$. The actual dissipation, given by Eq.~(\ref{e:BHLandauer}), is the smallest that does not permit such violations.

Notice that if one accounted for material accreting onto the black hole from the surrounding radiation bath, the average time needed for the black hole to lose one bit would be greater, but the black hole energy loss would still be the same, given by Eq.~(\ref{e:BHLandauer}). Of course on the timescale of individual bits, mass loss and gain will be stochastic.

Our conclusion is that a black hole carries out the erasure of its internal information at the most efficient dissipation level that is consistent with the Landauer Principle. This is our main result, which we term the {\it `Black hole saturation of the Landauer bound'}. 

Whether or not the information is actually erased entirely or is carried away with the emitted energy is a topic of longstanding debate in the gravity literature \cite{Almheiri,Raju}. For instance Zurek \cite{Zurek} (see also Page's response \cite{Page}, and his later work on the topic \cite{Page2}) found the cumulative entropy of the emitted radiation to be about $4/3$ that of the original black hole in the case of evaporation into a vacuum. Aghapour and Hajian \cite{AH} subsequently pointed out that the $4/3$ factor arises from general thermodynamic considerations not specific to black holes. 
%%%%%%% HERE FOR NEW CONTENT %%%%%%%
In Ref.~\cite{domenica-cortes-liddle-24} we show the $4/3$ factor found by Zurek in Ref.~\cite{Zurek} is a result of a modelling choice (emission onto a vacuum), and find that the numerical value distinguishing the emitted radiation from the original black hole equivalent depends on the choice of the space-time geometry that the radiation from the black hole is projected onto. %Zurek chose a flat, isotropic space geometry from which the factor of 4/3's derives, because we was using the traditional black body radition theory. See derivation of the Planck formula of waves in a box.
%%%%%%%%%%%%%%%%%%%%%%%%%%%%%%

Nevertheless, for our purposes here, we are applying the Landauer Principle only to the energy dissipation of the black hole in losing its information, and do not need to have an opinion on whether the entropy acquired by the surroundings equals or exceeds the lost bit, though this may ultimately determine the reversibility or irreversibility of the evaporation process.

Finally, we comment that astrophysical black holes are embedded in the 2.725 Kelvin cosmic microwave background, which exceeds the Hawking temperature for black hole masses greater than about that of the Moon (a few times $10^{22}$ kg). Absorption thus dominates over evaporation for all the expected astrophysical black holes, except for putative primordial black holes that may have formed in the early Universe \cite{PBH}.

\section{Existing literature}

Given the centrality of information in the discussions of black hole thermodynamics, it is somewhat surprising that there exists only a relatively small number of works directly connecting the former to algorithmic information and specifically to the Landauer Principle.\footnote{A different process also named after Landauer, namely Landauer Transport, has been deployed by various authors to discuss Hawking evaporation, starting with Ref.~\cite{Nation}.} Moreover, those works that have been made entirely concern the information loss of the exterior when material is accreted into a black hole.

The earliest study is by Fuchs \cite{fuchs}, who sought to use Landauer to support that the constant of the Bekenstein--Hawking entropy formula is one quarter of the event horizon area. There is no connection made to the Hawking evaporation process. A refined version of this set-up was then described by Song and Winstanley \cite{Song}.\footnote{That article took a rather surprising 8 years to pass from arXiv preprint to journal publication, despite the two versions being essentially identical.}. They considered a quantum system outside a black hole in thermal equilibrium with the Hawking radiation being emitted from it. Some entropy is allowed to fall into the black hole under the Landauer constraint and is shown to be compatible with the (generalised) second law of black hole thermodynamics.

The closest antecedent to our work is by Kim, Lee, and Lee \cite{KimLeeLee}. Their direction is to {\it assume} that black holes are the most efficient possible erasers of information, meaning erasing of information from their surroundings through accretion at the Landauer limit. This is then used to derive the black hole mass--entropy relation. Their Eq.~(4) matches our Eq.~(\ref{e:BHLandauer}) but the interpretation is different not only by the assumption above, but as the direction of mass and information transfer is into the black hole. 

Other work by these authors \cite{LeeLeeKim} sought a theory of dark energy from `forgetting', possibly akin to the information dark energy model of Gough motivated by the Landauer Principle \cite{Gough}.

More recently, Menin \cite{Menin} has sought to connect the Bekenstein bound \cite{Bekensteinbound}, limiting the amount of information that can be stored within a given physical volume, and the Landauer Principle, though the numerical results are specific to the supermassive black hole first imaged by the Event Horizon Telescope \cite{EHT-M87}. Denis \cite{Denis} compiles a number of results concerning black hole entropy and the Landauer Principle using entropic information, seeking to address the black hole information paradox.

\section{Extension to Barrow entropy}

An interesting extension to this network of relations is the Barrow entropy \cite{Barrow}, that may arise in place of the usual Bekenstein entropy due to a fractal event horizon structure. Abreu \cite{Abreu} has initiated investigation of this case via the Landauer Principle, though again focusing only on information lost into the black hole by the surroundings.

The simplest version of Barrow entropy takes a constant parameter $\Delta$ to run between zero (the usual case) and unity (maximally fractal, with the event horizon area scaling as if it were a volume) with entropy given by\footnote{We have included all the fundamental constants explicitly, unlike the cited papers.}
\begin{equation}
\label{e:BHentBarrow}
 \frac{S_{\rm BH}}{k_{\rm B}} =  \left( \frac{4 \pi G}{c \hbar} \,  M^2 \right)^{1+\Delta/2}\,.
\end{equation}
The thermodynamic identity, assuming constant volume, yields the corresponding temperature as \cite{Abreu} 
\begin{equation}
\label{e:tempdef}
\frac{1}{T_{\rm BH}} \equiv \frac{1}{c^2} \left.\frac{\delta S}{\delta M}\right|_V = \frac{k_{\rm B}}{c^2} (2+\Delta) \left(\frac{4 \pi G}{c\hbar}\right)^{1+\Delta/2} M^{1+\Delta} \,.
\end{equation}
Reassuringly, this reproduces the usual Hawking formula Eq.~(\ref{e:TBH}) for $\Delta = 0$. Nevertheless, more detailed analysis and some criticisms of the general validity of this simple approach, far beyond the scope of our present short article, can be found in Refs.~\cite{Abreu2,Saridakis,Nojiri,LuGennaroOng}. We will proceed assuming its validity.

Barrow \cite{Barrow} instead assumed that the black hole temperature remains proportional to $1/M$ despite the change in area, leading to a shorter black hole lifetime as the radiating area is greater. But this does not appear to be consistent with normal thermodynamic relationships, and our analysis suggests a longer lifetime as $\Delta$ is increased, since the suppression of the temperature dominates the greater area in the emission law 
\begin{equation}
{\rm Power} \propto {\rm Area} \times T_{\rm BH}^4 \,.   
\end{equation}
The implications of this for constraints on primordial black holes deserve to be investigated. 

Turning back to the Landauer Principle, a quick calculation following the one we did above affirms that the energy loss of the black hole per bit remains $\ln 2 \, k_{\rm B} T_{\rm BH}$ regardless of the value of $\Delta$. Hence the emission remains at the Landauer limit when considering the Barrow entropy. 

This result is due to the assumption that the radiation is emitted at the thermodynamic temperature, given by the first equality of Eq.~(\ref{e:tempdef}), from which the Landauer result follows directly. It might thus reasonably be argued that we have, to significant extent, put in the result by hand. Nevertheless, there is useful content in the statement because any other temperature relations that might be deployed (see e.g.\ Ref.~\cite{Nojiri}) would no longer saturate the Landauer limit. Some candidates might be considered ruled out were they to violate Landauer---as they would potentially enable Second Law violations---while those that exceed the Landauer limit may imply fundamental time irreversibility in the processes under consideration.

For instance, Barrow's supposition in Ref.~\cite{Barrow} was that the temperature continues to be given by Eq.~(\ref{e:TBH}), rather than by Eq.~(\ref{e:tempdef}). Since the former temperature is bigger (the ratio is about $M^\Delta$ in Planck units), the mass--energy loss for one bit would appear to be less than the corresponding Landauer limit, violating normal thermodynamical considerations and reinforcing Eq.~(\ref{e:tempdef}) as the more appropriate expression (see also Ref.~\cite{Nojiri}, and the application of Jacobson's \cite{Jacobson} thermodynamic approach to gravity theories in Ref.~\cite{LuGennaroOng}).

%\section{Quick note on boson stars - probably not relevant}

%Liebling, S.L., Palenzuela, C. Dynamical Boson Stars. Living Rev. Relativ. 15, 6 (2012). https://doi.org/10.12942/lrr-2012-6

%For a massive non-interacting scalar field of mass $\mu$ the maximum mass, a so-called boson star, is the Kaup mass
%\begin{equation}
%    M \simeq 0.633 \frac{c\hbar}{G\mu} \,.
%\end{equation}
%which has particle number
%\begin{equation}
%    N \simeq 0.653 \frac{c\hbar}{G\mu^2} \,.
%\end{equation}
%so that the gravitational binding energy is a few percent.

\section{Conclusions}

In this article we have demonstrated that the Hawking evaporation process \cite{Hawking} saturates the Landauer Principle~\cite{Landauer}. We termed this {\it `Black hole saturation of the Landauer bound'}, and it implies that an evaporating black hole loses its information as efficiently as is possible. While simple both conceptually and in calculation, we have not been able to find this result in the existing literature. It provides a link between black hole thermodynamics and algorithmic information theory. This link is related to, but distinct from, existing results on information loss to surroundings during accretion of material/information into a black hole, including the well-known Bekenstein bound \cite{Bekensteinbound,Casini}. We have also shown that this saturation of the Landauer bound holds  in the simplest incarnation of the Barrow entropy \cite{Barrow}. 

~\\   % To separate acknowldegements from above paragraph

\begin{acknowledgments}
We thank Paul Gough for introducing us to the Landauer Principle, Chris Fuchs for supplying the full text of Ref.~\cite{fuchs}, Adam Brown for important observations on the environment into which the radiation takes place, and Domenica Garzon, Shel Kaphan, Ken Matusow, Mark Neyrinck, Yen Chin Ong, and Tim Palmer for discussions and comments. This work was supported by the Funda\c{c}\~{a}o para a Ci\^encia e a Tecnologia (FCT) through the research grants UIDB/04434/2020 and UIDP/04434/2020. M.C.\ acknowledges support from the FCT through grant SFRH/BPD/111010/2015 and the Investigador FCT Contract No.\ CEECIND/02581/2018 and POPH/FSE (EC). A.R.L.\ acknowledges support from the FCT through the Investigador FCT Contract No.\ CEECIND/02854/2017 and POPH/FSE (EC). M.C.\ and A.R.L.\ were supported by the FCT through the research project EXPL/FIS-AST/1418/2021. 
\end{acknowledgments}

%\appendix

%\section{Appendixes}

%\nocite{*}

\end{document}